\documentclass[12pt, preprint]{aastex}
\usepackage{amssymb}

\shorttitle{De Buizer, Pi\~{n}a, \& Telesco}
\shortauthors{Mid-IR Observations of NGC 6334I}

\begin{document}

\title{Mid-Infrared Imaging of NGC 6334 I}
\author{James M. De Buizer\altaffilmark{1,2}, James T. Radomski %
\altaffilmark{3}, Robert K. Pi\~{n}a\altaffilmark{2,3}, and Charles M.
Telesco\altaffilmark{3}}

\slugcomment{Accepted for publication by the Astrophysical Journal} 

\altaffiltext{1}{Cerro Tololo Inter-American Observatory, National Optical Astronomy
Observatory, Casilla 603, La Serena, Chile.
CTIO is operated by AURA, Inc.\ under contract to the National Science
Foundation.} \altaffiltext{2}{Visiting Astronomer, W.M. Keck Observatory} 
\altaffiltext{3}{Department of Astronomy, University of
Florida, Gainesville, FL 32611}

\begin{abstract}
We present high-resolution ($<$0\farcs5) mid-infrared Keck II
images of individual sources in the central region of NGC 6334 I. We compare
these images to images at a variety of other wavelengths from the near
infrared to cm radio continuum and speculate on the nature of the NGC 6334 I
sources. We assert that the cometary shape of the UCHII region here, NGC
6334 F, is due to a champagne-like flow from a source on the edge of a
molecular clump and not a due to a bow shock caused by the supersonic motion
of the UCHII region through the interstellar medium. The mid-infrared
emission in concentrated into an arc of dust that define the boundary
between the UCHII region and the molecular clump. This dust arc contains a
majority of the masers in the region. We discuss the nature of the four
near-infrared sources associated with IRS-I 1, and suggest that one of the
sources, IRS1E, is responsible for the heating and ionizing of the UCHII
region and the mid-infrared dust arc. Infrared source IRS-I 2, which has
been thought to be a circumstellar disk associated with a linear
distribution of methanol masers, is found not to be directly coincident with
the masers and elongated at a much different position angle. IRS-I 3 is
found to be a extended source of mid-infrared emission coming from a cluster
of young dusty sources seen in the near-infrared.
\end{abstract}


\section{Introduction}

NGC 6334 is a parsec long train of rich molecular clouds and HII regions
located at galactic coordinates \textit{l}=351\arcdeg, \textit{b}=0.\arcdeg%
7. The complex lies at a distance of 1.74 kpc from the Sun \citep{Nec78},
parallel to and located in the Carina-Sagittarius spiral arm. It is the site
of possibly the largest number of recently formed OB stars observed in the
Galaxy, which may have been triggered by the recent passage of a spiral
density wave \citep{hg83}.

NGC 6334 was first discovered in the far-infrared by \citet{em73}. Later
observations in the far-infrared by \citet{mcb79}, revealed six centers of
emission. They were labeled by increasing southern declination using Roman
numerals I-VI. Our observations were of NGC 6334 I, the northernmost
far-infrared region of NGC 6334, and the site of a well-studied ultracompact
HII region, NGC 6334 F. Though heavily obscured at visual wavelengths, NGC
6334 I is the center of a wealth of activity in the infrared, millimeter,
and radio, as well as the site of many molecular sources and masers. Over
the decades, many authors have studied this region of NGC 6334, and each, it
seems, used nomenclature of their own to describe it. NGC 6334 I is a large
region that is identified with several significant sources summarized by %
\citet{KDJ99}. We will use the convention NGC 6334 F from the radio
continuum observations of \citet{rcm82} to describe the UCHII region we
observed in the mid-infrared. The HII region is clearly cometary shaped in
the radio \citep{rcm82,drdg95,enm96}, millimeter \citep{ckrdh97}, and
mid-infrared \citep{DPT00,ptf98}, with its head pointing to the northwest
and the tail running to the southeast. The peak of the UCHII region lies
near the infrared source IRS-I 1 of \citet{bn74}, which has been presumed to
be the ionizing source of the HII region. \citet{hg83} also find another
source $\sim $6$\arcsec$ to the northwest of IRS-I 1, designated IRS-I 2,
and yet another $\sim $18$\arcsec$ to the east, designated IRS-I 3.

This region is very complex and is the site of a wide variety of activity. A
near-infrared survey by \citet{t96} found an embedded young cluster of 93
sources associated with NGC 6334 I, all within a radius of $\sim $80$\arcsec$%
. This cluster, of which IRS-I 1, IRS-I 2, and IRS-I 3 are members, appears
to only contain stars earlier than B3-B4 according to \citet{t96}. In light
of the complexity of the NGC 6334 I area, interpretation of data is not an
easy task. In this paper we present high-resolution mid-infrared images of
the sources within NGC 6334 I. In \S 2 we will discuss the observations of
NGC 6334 I, and explain the data reduction process in \S 3. Interpretation
of our data and a discussion of the phenomenology of each source will be
presented in \S 4. Finally, in \S 5 we will present our conclusions.

\section{Instrumentation and Observations}

Observations of NGC 6334 I were carried out in May of 1998 and 1999 on the
Keck II 10-m telescope on Mauna Kea. Broadband $N$\ ($\lambda _{o}$=10.46 $%
\micron$, $\Delta \lambda $=5.1 $\micron$) and $IHW18$ (International Halley
Watch, $\lambda _{o}$=18.06 $\micron$, $\Delta \lambda $=1.7 $\micron$)
imaging was performed using OSCIR, the University of Florida mid-infrared
camera/spectrometer. OSCIR is equipped with a 128$\times $128 pixel,
silicon/arsenic-doped blocked impurity band (Si:As BIB) array which is
optimized for 8-25 $\micron$ work.

At Keck II, OSCIR has a field of view of 8\arcsec$\times $8\arcsec, for a
scale of 0\farcs0616 pixel$^{-1}$. Sky and telescopic radiative offsets were
subtracted using a secondary chopping at 5 Hz, and by nodding the telescope
every 15 seconds. Frame times of 15 ms were used for all observations.
Images presented within this paper have total on-source integration times in
both filters of 120 seconds for IRS-I 1, 480 seconds for IRS-I 2, and 240
seconds for IRS-I 3. The standard star $\alpha $ Boo was observed at roughly
the same airmass ($\thicksim $1.7) as NGC 6334 I. It was used as a flux
calibrator, with flux densities taken to be 683 Jy at $N$ and 219 Jy at $%
IHW18$. Point-spread function (PSF) stars, were also imaged near the
position of NGC 6334 I, yielding a measured full width at half maximum
(FWHM) of 0\farcs33 at $N$ and 0\farcs41 at $IHW18$.

\section{Results and Data Reduction}

We have observed NGC 6334 I previously as part of a mid-infrared survey
using the Cerro Tololo Inter-American Observatory 4-m telescope \citep{DPT00}%
. The field of view at CTIO is large enough to encompass not only the UCHII
region, but also the regions where IRS-I 2 and IRS-I 3 reside (Figure 1). We
were able to image the whole central region of NGC 6334 I at both 10 and 18 $%
\micron$, though the extended emission from IRS-I 3 is slightly off field in
both filters. Because of the small field of view at Keck, we imaged the
regions containing IRS-I 1, IRS-I 2, and IRS-I 3 individually (Figure 1). %
\citet{ckrdh97} detect an additional source just southeast of IRS-I 2 at 7
mm. Not only was it not detected in the $J$, $H$, and $K$ images of %
\citet{per96}, we did not detect it in our images from CTIO at either 10 or
18 $\micron$ \citep{DPT00}$.$ Due to time constraints, we did not obtain
follow-up images centered on this 7 mm source at Keck.

\begin{figure}
\plotone{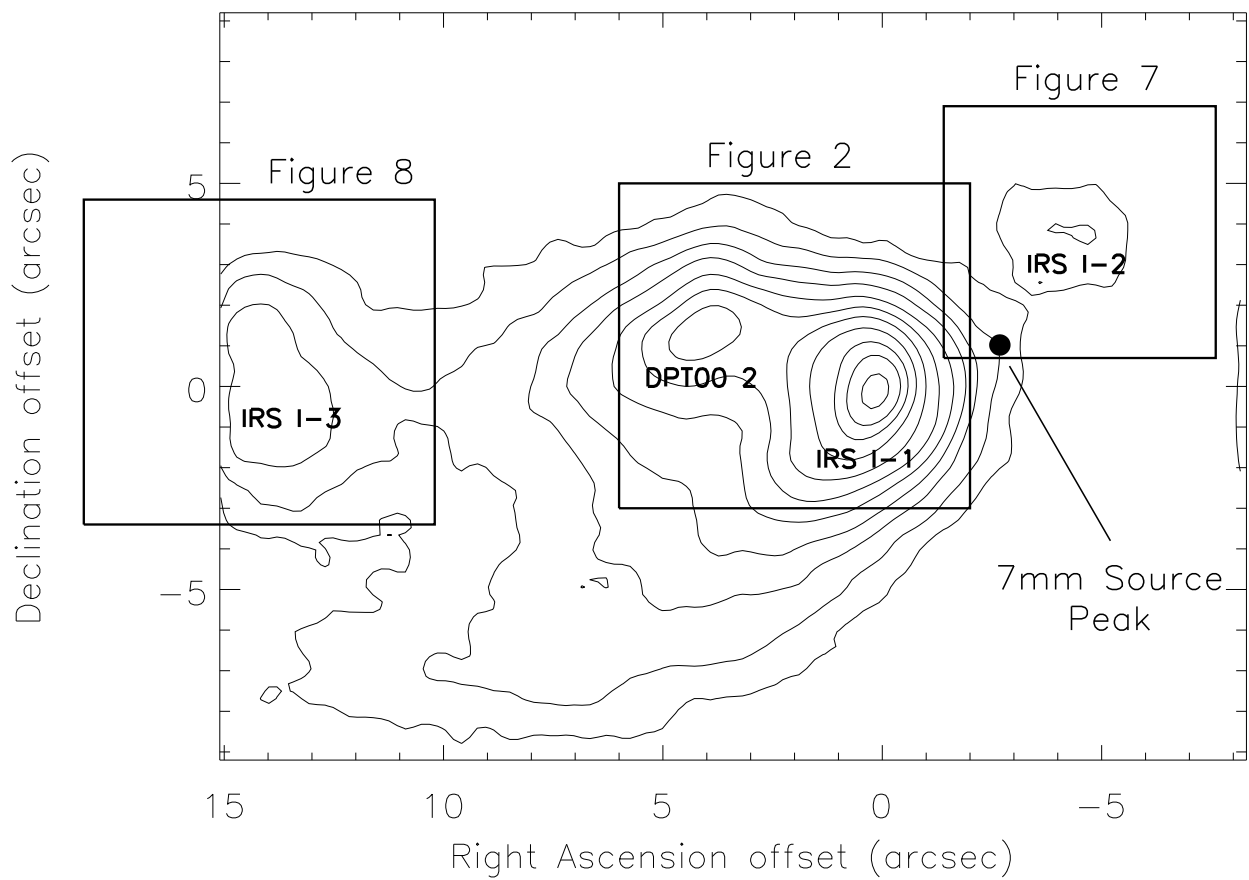}
\caption{The central region of NGC 6334 I at 18 {\micron}. 
This image was taken at the CTIO 4-m by De Buizer, Pi\~{n}a, \& Telesco (2000)
at a resolution of $\sim$1$\arcsec$. The individual fields
that were imaged at Keck are shown as boxes. At the center of this image is 
NGC 6334 F, a cometary UCHII region. It can be seen as two separate sources
in the mid-infrared, IRS-I 1 and DPT00 2. Also labeled is the location of the 
7 mm source seen by Carral et al. (1997). This source was not detected in our 
mid-infrared images. The coordinates of the origin are 
R.A.(J2000)=17$^h$20$^m$53\fs44 and  
Decl.(J2000)=-35$\arcdeg$47$\arcmin$02\farcs2.}
\end{figure}

\subsection{Flux Calibration}

Because of the large bandwidth of the filters, the observed fluxes must be
color corrected to account for the intrinsic source spectrum, the filter
transmission, and the atmospheric transmission. For the calibration stars,
the spectra were assumed to be a blackbody at the effective stellar
temperature. Color corrected monochromatic flux densities, dust color
temperatures, and optical depth values were obtained in a self-consistent
manner by iteratively performing a numerical integration on the product of
the Planck function, emissivity function (given by 1-e$^{-\tau _{\lambda }}$%
, where $\tau _{\lambda }$ is given by the extinction law of Mathis 1990),
filter transmission, solid angle subtended by the source, and model
atmospheric transmission through the filter bandpass. A detailed treatment
of this color-correction method is given in De Buizer (2000). As is often
the case with mid-infrared observations, the calibration factor (ratio of
accepted flux in Jy to analog-to-digital converter units per second per
pixel) derived from the standard star observations changed throughout the
course of the night due to changes in atmospheric conditions, but there
was little correlation with airmass. Therefore, airmass corrections were not
made to the observations. We can, however, estimate the absolute photometric
accuracy (i.e. the mean calibration value of the standard star observations
throughout the night divided by the standard deviation) associated with the
tabulated color corrected flux densities in Table 1 to be 6.8\% at 10 $%
\micron$\ and 9.7\% at 18 $\micron.$

\subsection{Temperature and Optical Depth Maps}

One advantage of acquiring images at two wavelengths at Keck II is that we
were able to construct dust color temperature (T) and emission optical depth
($\tau $) maps for each source. This was accomplished by first convolving
the 10 $\micron$ source images with the 18 $\micron$ image of a point-spread
function (PSF) star, and the 18 $\micron$ source images with the 10 $\micron$
PSF image. This step is important because artificial structures in the
temperature and optical depth maps can result from failure in having both
images at the same resolution. This convolution process gives an effective
resolution of the temperature and optical depth maps of 0\farcs53. The
relative alignment of the two images is also crucial to the values derived
for temperature and optical depth. We employed an automated registration
algorithm based on minimizing the sum of the squared residuals of the image
difference as a function of the relative offsets. This algorithm generates a
``chi-squared'' surface at integral x and y pixel offsets. The chi-squared
surface may then be interpolated to determine the location of the minimum to
a hundredth of a pixel. Best-fit alignments for the image sets were found in
this way for all sources.

Once an image set was spatially registered, the 10 and 18 $\micron$ flux
densities for each pixel were used to iteratively solve for emission optical
depth ($\tau $) at 10 $\micron$ and dust color temperature under the
assumption of blackbody emission. For these calculations we used the
relationship $\tau _{10\micron}=\tau _{18\micron}/1.69$ from the extinction
law of Mathis (1990). A cut-off was applied to both the 10 and 18 $\micron$
flux density maps at 3$\sigma $ above the background sky value. Temperature
and optical depth values were only calculated for those areas that were
above this cutoff at both wavelengths. The relative alignments of the 18 $%
\micron$ and 10 $\micron$ source images were then shifted by 4 pixels (0$%
\farcs$25) in various directions. It is unlikely that our registration of
the images is off by such a large amount, however this allows a test of the
robustness of the results concluded from the temperature and optical depth
maps. It was found that these shifts significantly changed the peak
temperature and optical depth values ($\pm 25\%$), however these shifts
created only slight changes ($<0\farcs3$) in the peak locations and overall
morphologies. Therefore, while there may be uncertainty in the absolute
values for these temperatures and optical depths, the maps are quite useful
in demonstrating the spatial trends of these properties for these sources.
The lower spatial scale color temperature and optical depth maps of Kraemer
et al. (1999) look very similar to our maps, indicating that the structures
in these maps are real.

\subsection{Astrometry}

Because of time constraints, careful astrometry was not performed.
Therefore, accurate astrometry of the images needed to be performed by
registering the mid-infrared images with other wavelength data. In the case
of \citet{DPT00}, the 5.0 cm radio continuum map of \citet{c97} was
registered with the mid-infrared images by aligning the radio and
mid-infrared morphologies and peaks. Even with the higher spatial resolution
images from Keck, is was found that the astrometry used in \citet{DPT00}
does indeed show the best morphological coincidences between the
mid-infrared images of NGC 6334 F and images at other wavelengths. We can
not be completely sure that this is the correct astrometry because the
morphologies may be wavelength dependent. However, this registration does
show surprisingly good morphological coincidences throughout a broad range
of wavelengths (6.2 cm, 5.0 cm, 3.5 cm , 2.0 cm, 1.3 cm, 2.2 $\micron$), as
we will discuss in\ more detail in \S 4.1. We therefore feel that the
absolute astrometry of our mid-infrared sources is good to $\leq $0\farcs5.

\subsection{Luminosity and Spectral Type}

Estimates of ZAMS stellar luminosities were made for each of the objects
based upon their mid-infrared color-corrected fluxes. These are presented in
Table 1. These values are based on the simplified assumption that all of the
luminosity seen at the mid-infrared wavelengths is dust-reprocessed stellar
radiation, and so is indicative of the true bolometric luminosity of the
stellar source itself. These mid-infrared luminosity estimates were computed
by integrating the Planck function from 1 to 600 $\micron$ at the derived
dust color temperature and optical depth for each source. This calculation
employs the above emissivity function and assumes emission into 4$\pi $ sr.
Using the tables of \citet{Doyon}, which are based on the stellar
atmospheric models of \citet{K79}, we then found the ZAMS spectral types
associated with those mid-infrared derived luminosities.

The main problems with this method of deriving estimates to the bolometric
flux are 1) if the dust is anisotropically distributed around the source,
the derived luminosity would depend on this dust distribution because some
of the stellar flux will escape unprocessed through the unobstructed
regions; 2) heavy obscuration could lead to non-negligible reprocessing by
dust of the mid-infrared photons into far-infrared photons; 3) dust is in
competition with gas for the short wavelength photons, which ionize the gas
and produce UCHII regions. All of these processes would lead to
underestimates of the bolometric luminosities from mid-infrared fluxes,
however it is hard to quantify exactly how each contribute. For these
reasons we believe that the derived bolometric luminosities represent good
lowers limit to the true bolometric luminosities.

As we will discuss later in the paper, some of these sources are believed
not to be centrally heated. Therefore, the derived ZAMS spectral types in
reality will not apply, and the luminosity given in Table 1 is a better
indication of the infrared luminosity of the source, rather than the
bolometric luminosity.

\section{Discussion of Individual Sources}

\subsection{NGC 6334 F (IRS-I 1 and DPT00 2)}

Mid-infrared observations of the UCHII region NGC 6334 F have revealed that
it is composed of two sources \citep{DPT00, KDJ99}. Our brightest
mid-infrared peak (Figure 2) is apparently the same as the peak of the
cometary UCHII\ region\ seen in the radio maps of NGC 6334 F. Another
mid-infrared source lies 4$\arcsec$ north-east of the mid-infrared peak, and
is elongated in its thermal dust distribution. This source has been
designated G351.42+0.64:DPT00 2 by \citet{DPT00}, and is also referred to as
NGC 6334:MFSW I:KDJ 4 from \citet{KDJ99}. Throughout this paper we will
refer to this source as DPT00 2 (Figure 1).

\begin{figure}
\plotone{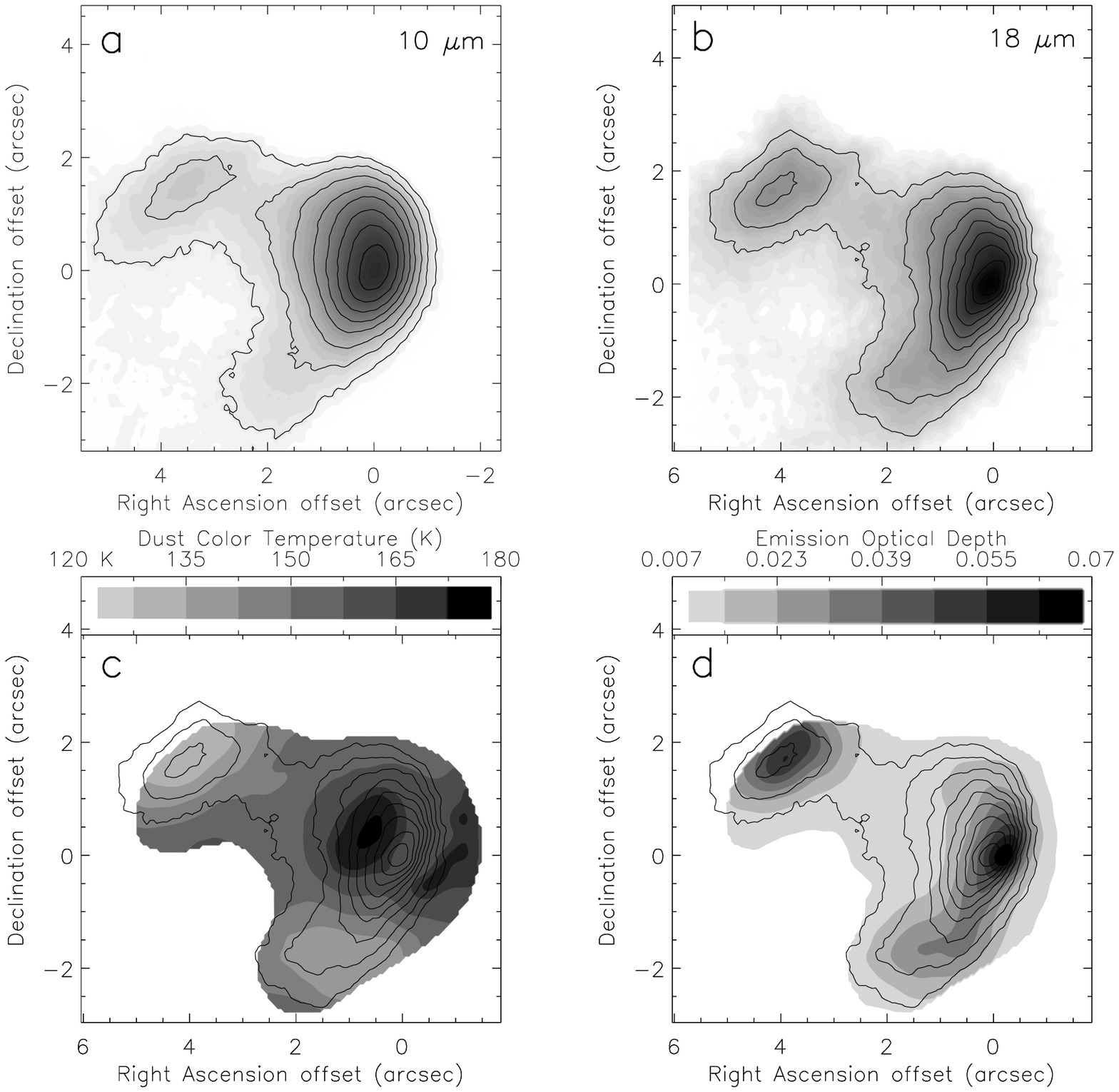}
\caption{Keck data for NGC 6334 F. Panels a and b show the 
OSCIR images in grayscale at 10 and 18 {\micron}, respectively, convolved with a 
Gaussian kernel of FWHM 2 pixels. Contours are added 
for emphasis. In panel a, contour levels are 3, 4, 5, 6, 9, 12, 16, and 22 Jy arcsec$^{-2}$. 
In panel b, contour levels are 6.5, 10, 15, 20, 30, 40, 50, 60, and 80 Jy arcsec$^{-2}$.
Panel c shows the color temperature map of the region, and panel d 
shows the emission optical depth map, both with the 18 {\micron} contours 
overlaid. The origin in each panel is the same as in Figure 1.}
\end{figure}

\subsubsection{Temperature and Optical Depth Maps}

In Figure 2d, one sees that the 18 $\micron$ contours of IRS-I 1 delineate
the optical depth contours (grayscale) fairly well. Looking at an overlay of
the dust color temperature and the 18 $\micron$ map (Figure 2c), the
temperature peak is offset from the 18 $\micron$ peak. Both of these
observations indicate that the mid-infrared emission is bright there simply
because it is optically thin in the mid-infrared and we are seeing through
more material in the line of sight. This is a ridge of material that could
have be swept up dust from the expanding shock front of the UCHII region.
This leads to the conclusion that the mid-infrared peak may not be
delineating the location of the exciting stellar source for the UCHII region.

\subsubsection{Radio Continuum}

The UCHII region of NGC6334F has been extensively observed at radio
wavelengths. Due to similar spatial resolution (FWHM$\thicksim $0.5$\arcsec$%
) and morphology, we used the 3.5 cm radio maps of \citet{ckrmca02} to
achieve accurate astrometry of our mid-infrared maps (Figure 3a). Using this
astrometry, we also checked the mid-infrared morphology against other radio
contour maps at similar resolution: the 5 cm (FWHM$\thicksim $2$\arcsec$) of %
\citet{c97}, the 3.6 cm (FWHM$\thicksim $1.2$\arcsec$) of \citet{enm96}, and
the 2.0 cm map (FWHM$\thicksim $1.0$\arcsec$) of \citet{drdg95} (Figure 3b).
In all cases, the radio contours corresponded closely to those of the
mid-infrared around the UCHII region peak, adding confidence to the idea
that the radio and mid-infrared emission are coming from the same location.
Comparing the radio maps with the mid-infrared images shows that radio
emission comes not only from the location of the peak of the UCHII\ region,
but from DPT00 2 as well. The radio continuum emission seems to trace the
southern edge of DPT00 2 where it faces the peak of the UCHII\ region,
indicating that it may be externally ionized. The color temperature maps are
consistent with this scenario. The southern edge of DPT00 2 is also the
hottest part of the source, and therefore may not be internally heated
(Figure 2c).

\begin{figure}
\plotone{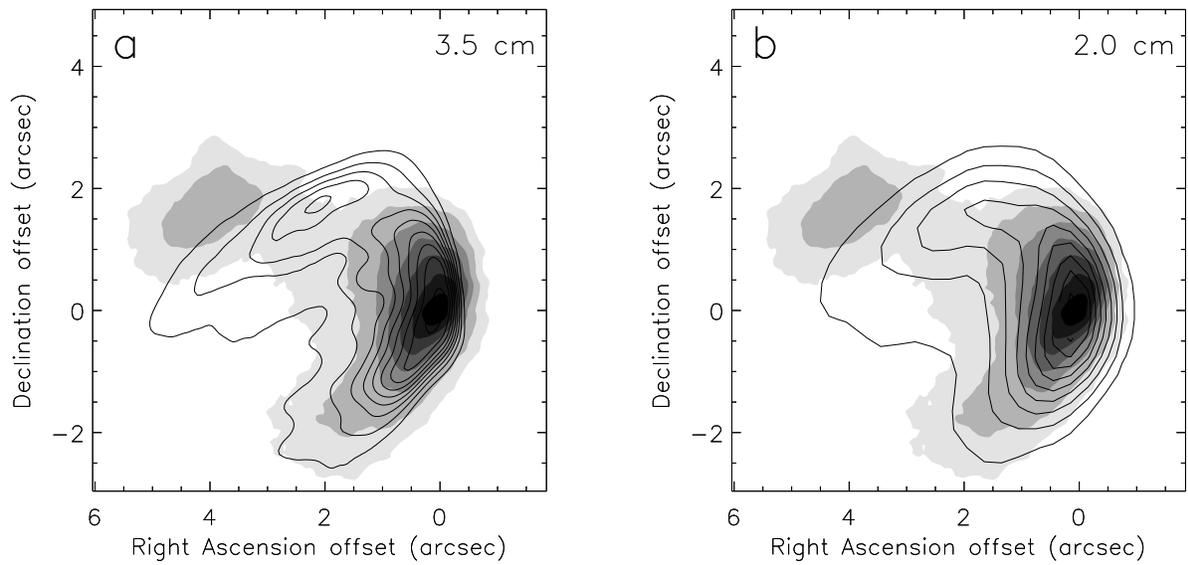}
\caption{Comparisons of the 18 {\micron} images (filled contours) 
with radio continuum images (contours) demonstrate that the contours of the radio 
continuum and mid-infrared are correlated for IRS-I 1. Panel a shows the 3.5 cm radio continuum images of Carral et 
al. (2002) registered with respect to the mid-infrared image. By registering the mid-infrared 
images with respect to the the radio continuum images, the mid-infrared absolute 
astrometry was determined. Panel b 
shows the registration between the 2.0 cm maps of De Pree et al. (1996) and the 
mid-infrared.  The origin in both panels is the same as in Figure 1.}
\end{figure}

\subsubsection{Comparisons to the Observations of Harvey and Gatley (1983)}

The peak in the UCHII region at 18 $\micron$ apparently corresponds to the
IRS-I 1 peak of \citet{hg83}. Overlays between our 18 $\micron$ mid-infrared
maps and the 20 $\micron$ mid-infrared maps of \citet{hg83} show similar
morphology with two peaks with the same separations as IRS-I 1 and IRS-I 2
(Figure 4a). The 20 $\micron$ maps of \citet{hg83} show extension in the
direction of DPT00\ 2, but do not resolve the source. Using the astrometry
above, we find that the absolute coordinates of the 20 $\micron$ peaks of
IRS-I 1 and IRS-I 2 from \citet{hg83} are both in error by approximately 1%
\farcs5 in the southern direction. This is entirely plausible, given that
these observations were performed by scanning the area with a photometer
with a 4$\arcsec$ beam, though the quoted positional accuracy is 1$\arcsec$.
Assuming our astrometry is correct and using the coordinates of the peak of
the radio continuum from \citet{ckrmca02}, the mid-infrared peak of IRS-I 1
is at R.A.(J2000)=17$^{h}$20$^{m}$53\fs44, Decl.(J2000)=-35$\arcdeg$47$%
\arcmin$02\farcs16. This changes the mid-infrared coordinates for the other
sources as well. These new coordinates are given in Table 1. 

\begin{figure}
\plotone{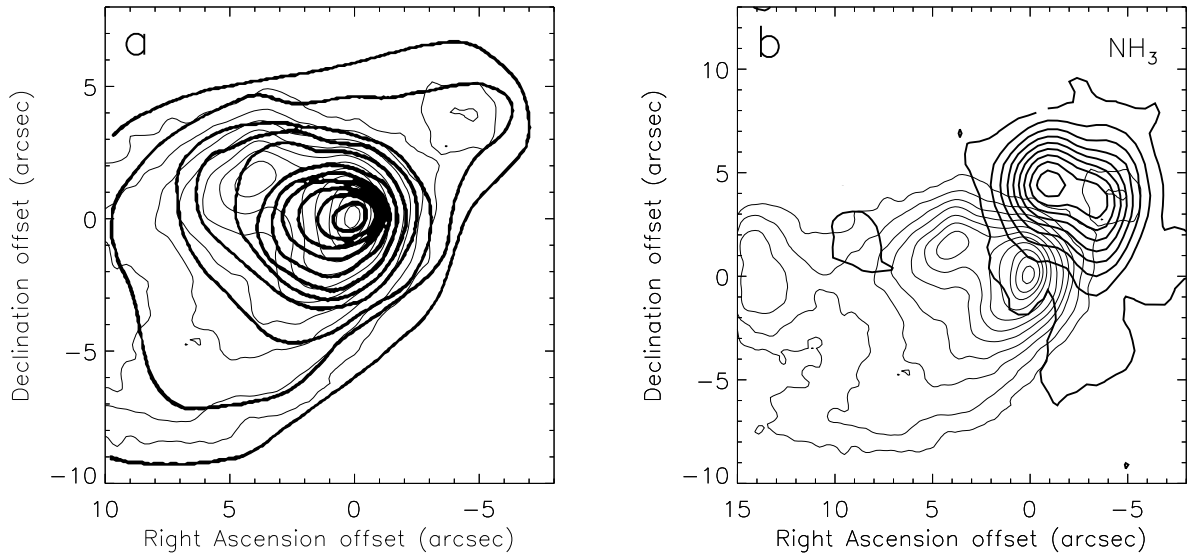}
\caption{Comparisons of the 18 {\micron} emission from NGC 6334 I 
with other large-scale emission. Panel a shows the 20 {\micron } contours (thick) of 
Harvey and Gatley (1983) compared to the 18 {\micron} contours (thin) from De Buizer, Pi\~{n}a, \& Telesco (2000). A very good match is found, leading us to a revision in the mid-infrared 
coordinates of Harvey and Gatley (1983). Panel b shows the integrated ammonia map 
(thick contours) of Kraemer et al. (1999) registered with respect to the  
mid-infrared images from De Buizer, Pi\~{n}a, \& Telesco (2000, thin contours) using the new mid-infrared coordinates. The ammonia 
appears to bound the UCHII region on to the west, and may be responsible for the 
champagne-like flow manifesting itself as the cometary shaped UCHII region. The 
origin in both panels is the same as in Figure 1.}
\end{figure}

\subsubsection{Ammonia Distribution}

Ammonia observations by \citet{KJ95} show the UCHII region to be bounded by
ammonia (3,3) emission to the west. This emission is a dense gas indicator,
as is CS 7$\rightarrow $6 which also peaks west of the UCHII region %
\citep{KDJ99}. It appears therefore that the star responsible for ionizing
the UCHII region exists on the edge of this density enhancement, and this
gradient in the surrounding medium is what has led to the shape of the UCHII
region. These observations seem to indicate the cometary appearance is due
to a champagne-like flow, rather than a bow-shock caused by supersonic
motion through the interstellar medium.

In Figure 7 of \citet{KDJ99}, they show an overlay of their mid-infrared and
ammonia observations by registering them using the coordinates of %
\citet{hg83}, which we believe to be in error. We present in Figure 4b the
correctly registered integrated ammonia map of \citet{KDJ99} and our
mid-infrared map of the region. Because the radio continuum and ammonia maps
were taken at the same time, they have extremely accurate relative
astrometry. We overplotted the integrated ammonia in Figure 4b by first
registering the radio continuum peak of \citet{KDJ99} with our mid-infrared
peak for IRS-I 1. The integrated ammonia map presented in Figure 4b is
mostly ammonia emission but also contains ammonia seen in absorption.
The weakest contour of ammonia in the integrated map appears to wrap around
the mid-infrared peak in Figure 4b. This ammonia component is associated
with the mid-infrared (and radio continuum) peak and is actually seen in
aborption by Kraemer and Jackson (1995). The rest of the integrated map
shows ammonia emission, bounded by IRS-I 1 to the east and IRS-I 2 lies just
to the west of the secondary peak. The ammonia emission follows along the
eastern edge of the UCHII region, extending to the south.

\subsubsection{Masers}

The maser emission in this region is concentrated near IRS-I 1 (Figure 5).
This area is marked by several molecular maser species:\ methanol \citep{E96}%
, water \citep{fc89}, and hydroxyl \citep{gm87}. A\ majority of the masers
seem to be on the sharp western edge of the UCHII region which is bounded by
the ammonia emission (Figures 5a and 5b). This density enhanced side of the
UCHII region will also be a location where the expansion of the ionized
material will first impact as a shock front. The density and energetics of
such a region would be suitable for creating and sustaining maser emission.
If this is the case, the masers that exist on this sharp boundary may be
shock-induced.

\begin{figure}
\includegraphics[height=17cm,angle=90]{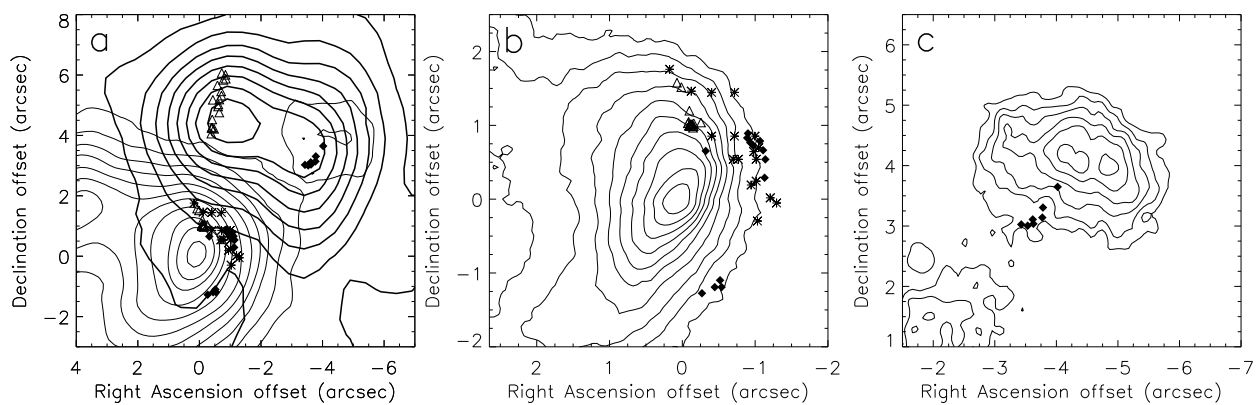}
\caption{The water (triangles), hydroxyl (stars), and methanol (filled 
diamonds) masers in the NGC 6334 I region. Panel a shows that a majority of the 
masers are associated with the sharp boundary between the 18 {\micron} emission (thin 
contours) of IRS-I 1 and the ammonia emission (thick contours). This 18 {\micron} image 
is from the CTIO 4-m (De Buizer, Pi\~{n}a, \& Telesco 2000). A string of water 
masers may be associated with the brightest peak of the ammonia distribution, and a 
string of methanol masers may be associated with the secondary peak. Panel b 
zooms in closer to IRS-I 1 to show that the masers are all excited along the sharp western 
edge of the source. Panel c shows that the methanol masers that were thought to be 
associated with IRS-I 2 are not coincident with the mid-infrared source peak. 
The origin in each panel is the same as in Figure 1.}
\end{figure}

There is a long string of water masers that are offset to the north of the
mid-infrared emission from the IRS-I 1. Strings of water masers can be
interpreted as coming from and delineating outflow (e.g. Claussen et al.
1997). However, water masers are also known to exist coincident with ``hot
cores'' as seen from their molecular emission \citep{CCHWK94}. Observations
of G9.62-0.19 \citep{HC96} show a string of water masers emanating radially
from a UCHII region center. However, the ammonia observations of %
\citet{CCHWK94} show a peak at the same position as the water masers, and it
has been proposed that these water masers are delineating the location of a
hot core. Likewise, given the corrected astrometry set forth in this paper,
the primary ammonia peak north of IRS-I 1 is coincident with the water maser
string. It is plausible that these water masers are delineating the site of
a hot core.

There is also a group of methanol masers that are offset northwest of IRS-I
1, that are thought to be associated with IRS-I 2 (Figure 5c). Our
astrometry shows that they are not exactly coincident with this mid-infrared
source, and are close to the secondary ammonia peak (Figure 5a). These
methanol masers may be delineating a second hot core in the ammonia. Both of
these maser strings may delineate the location of hot cores that are too
cool and/or embedded to see in the thermal infrared.

\subsubsection{Mid- Infrared versus Near Infrared Sources}

\citet{per96} observed this area in the near infrared bands J(1.25 $\micron$%
), H(1.65 $\micron$) and K(2.2 $\micron$) at a resolution of 0\farcs9. These
observations revealed 4 sources within 3$\arcsec$ of the IRS-I 1 peak. The
near infrared sources are labelled IRS1E, IRS1W, IRS1SE, and IRS1SW (Figure
6). These sources are highly reddened and therefore believed not to be
foreground stars. \citet{per96} claims that the IRS1E source is coincident
with the 30 $\micron$ peak of \citet{hg83}. Overlaying the K image with our
mid-infrared images shows good morphological coincidences with both IRS-I 1
and IRS-I 3 (see \S 4.3), leading us to believe the relative astrometry
between the near-infrared and mid-infrared to be better than 0\farcs5. This
alignment places IRS1E peak approximately 1\farcs8 from the 30 $\micron$
peak and 0\farcs7 from our 18 $\micron$ peak.

Overlaying the K band image with the color temperature map generated from
our mid-infrared images we see a very important coincidence that once again
leads us to believe our astrometry is correct (Figure 6d). The color
temperature is peaked near the location of a near infrared source (IRS1E).
Also, there is a near infrared source seen at J and H, and extended at K,
coincident with the south-eastern portion of DPT00 2 (Figure 6c). We see in
the temperature map in Figure 6d that the hottest parts of DPT00 2 are on
the side facing the UCHII region. Since this region is hotter, it should be
more readily seen in the near infrared. These coincidences of the hotter
regions to near infrared emission seem to confirm our relative astrometry.
The coincidence of IRS1E to the peak in the color temperature map also leads
us to speculate that the near infrared source IRS1E may be responsible for
the central heating and ionizing of the the UCHII region.

\begin{figure}
\plotone{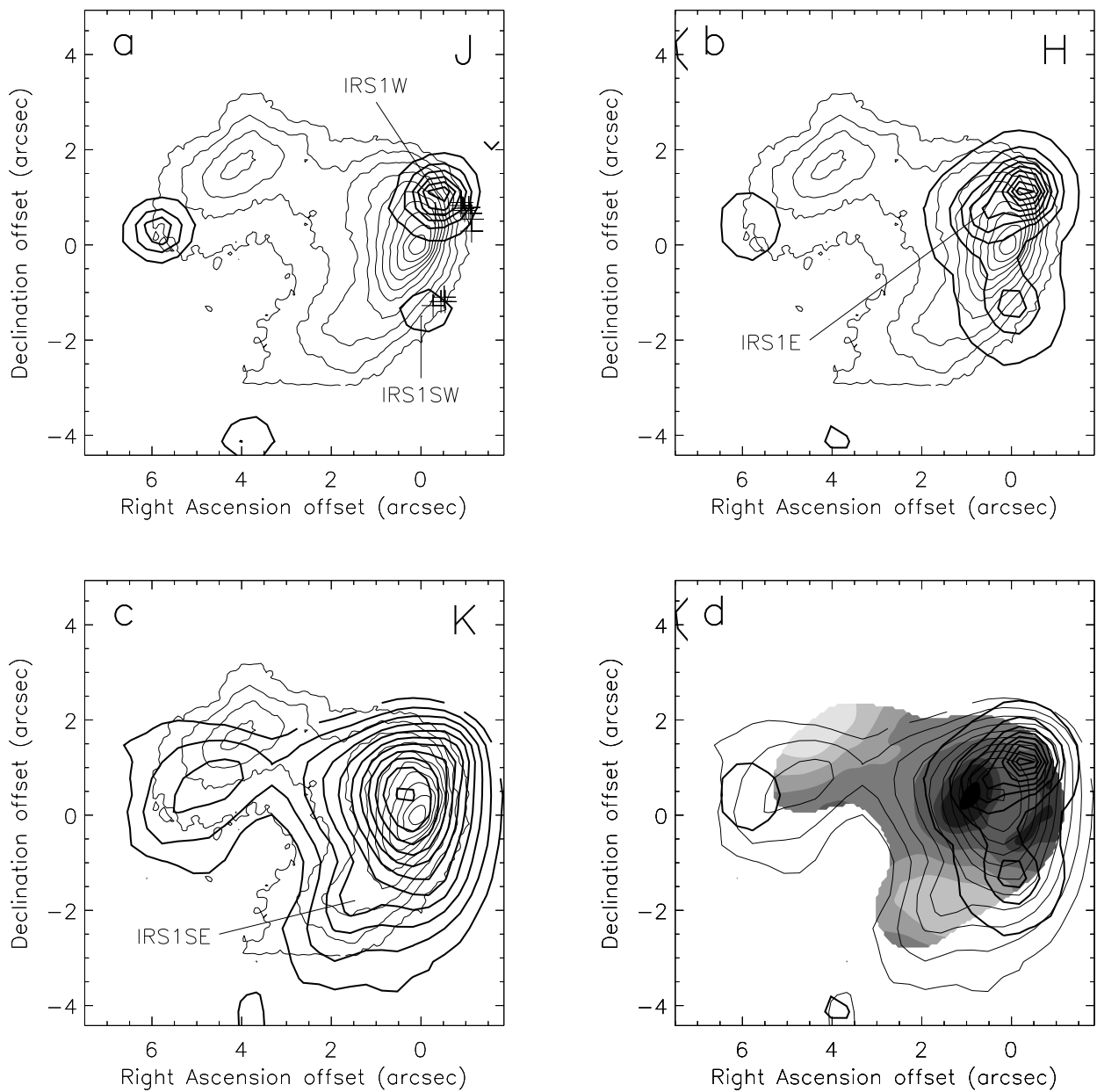}
\caption{The near infrared sources in the region near IRS-I 1 from Persi 
et al. (1996). Panel a, b and c, show the near-infrared (thick contours) J, H, and K images, 
respectively, overplotting the 18 {\micron} Keck contours (thin). Labels are the names 
of the near-infrared sources given by Persi et al. (1996). The methanol masers are plotted in 
panel a as crosses. They appear to be associated with two near infrared sources. 
Panel d shows the color temperature map (filled contours) overlaid with the K emission 
(thin contours) and H emission (thick contours).  The origin in each panel is the same 
as in Figure 1.}
\end{figure}

The nature of IRS1W and IRS1SW can also be speculated from the color
temperature and optical depth maps. By looking at the near-infrared images
of \citet{per96} in Figure 6, we see that as one views this region at
shorter and shorter wavelengths the near-emission comes only from the
sources within the mid-infrared arc of emission from IRS-I 1. Since the
optical depth is largest at the mid-infrared peak, this is where there is a
higher concentration of cooler material. At J and H, IRS1W and IRS1SW lie
just north and just south of the mid-infrared peak (Figures 6a and 6b). The
near-infrared emission is coming from areas where thermal dust emission is
located, but not at its densest and brightest parts. Furthermore, both of
these sources are not \ temperature peaks in the color temperature map
(Figure 6d). If IRS1W and IRS1SW are not hot, there must be an alternate
reason why they are visible in the near infrared.

One scenario is that IRS1W and IRS1SW are visible in the near-infrared
because of shock excited emission \citep{D02}. These sources are most
prominent at J and H, and both of these filter bandpasses encompass many
lines of [FeII]. These spectral lines are shock indicators \citep{MCH84},
two of the strongest being the lines at 1.26 and 1.64 $\micron$ (roughly the
central wavelengths of the J and H filters). \citet{B98} observed a diffuse
UCHII region like NGC 6334 F and find that the thermal dust emission and
narrow line [FeII] images are well matched spatially. As discussed above,
the shape of the UCHII region and the presence of masers on the sharp
western boundary imply that the masers may be shock excited. In this same
location there is a density enhancement of dust, as seen in the
mid-infrared, which may be swept up material from an expanding shock front
rather than circumstellar in origin. IRS1W is located on this dust ridge and
coincident with the majority of the masers here and is most prominent in the
J and H bands, which contain several lines that are shock indicators. For
all of these reasons, it seems plausible that IRS1W and IRS1SW are not
stellar sources themselves, but instead may be areas of shock excited
emission. The final proof of such a hypothesis would be to obtain
near-infrared spectra of these sources.

However, inconsistent this scenario is the near-infrared H$_{2}$ observation
shown in the article by \citet{per96} that does not seem to show any
shock-excited H$_{2}$ emission at these locations. \citet{G91} finds
that H$_{2}$ emission (a more widely used shock indicator) and [FeII] are
well correlated in their emission. Therefore, an alternative scenario could
be that IRS1W and IRS1SW are simply areas of reflected emission from the
less-extinguished parts of IRS-I 1.

As for the near-infrared source IRS1SE, we see no mid-infrared source or
peak at this location, but it is an area of diffuse mid-infrared emission.
It seems that IRS1SE is not a stellar source either, and that the
near-infrared emission may likely be coming from reflected or scattered
photons off of the tail of the UCHII region.

\subsection{IRS-I 2}

The Keck images of IRS-I 2 reveal it to be an elongated and low-surface
brightness source. The source is relatively low signal-to-noise (peak pixel
S/N$\sim 7$), but smoothing (3 pixel gaussian) shows three peaks in the
thermal emission along the direction of elongation (Figure 7). The central
and western peaks are comparable in brightness, however the eastern source
is fainter, especially at 18 $\micron$. Because the western peak can be
clearly seen at both 10 an 18 $\micron$, we chose this location as the
reference position for IRS-I 2 in Figure 7, and give the coordinates for
this location in Table 1. Overlays of the NGC 6334 I region from \citet{hg83}
and our 18 $\micron$ data show a good match in the peak locations,
confirming that the source seen at Keck is indeed IRS-I 2 (Figure 4a). \ 

\begin{figure}
\plotone{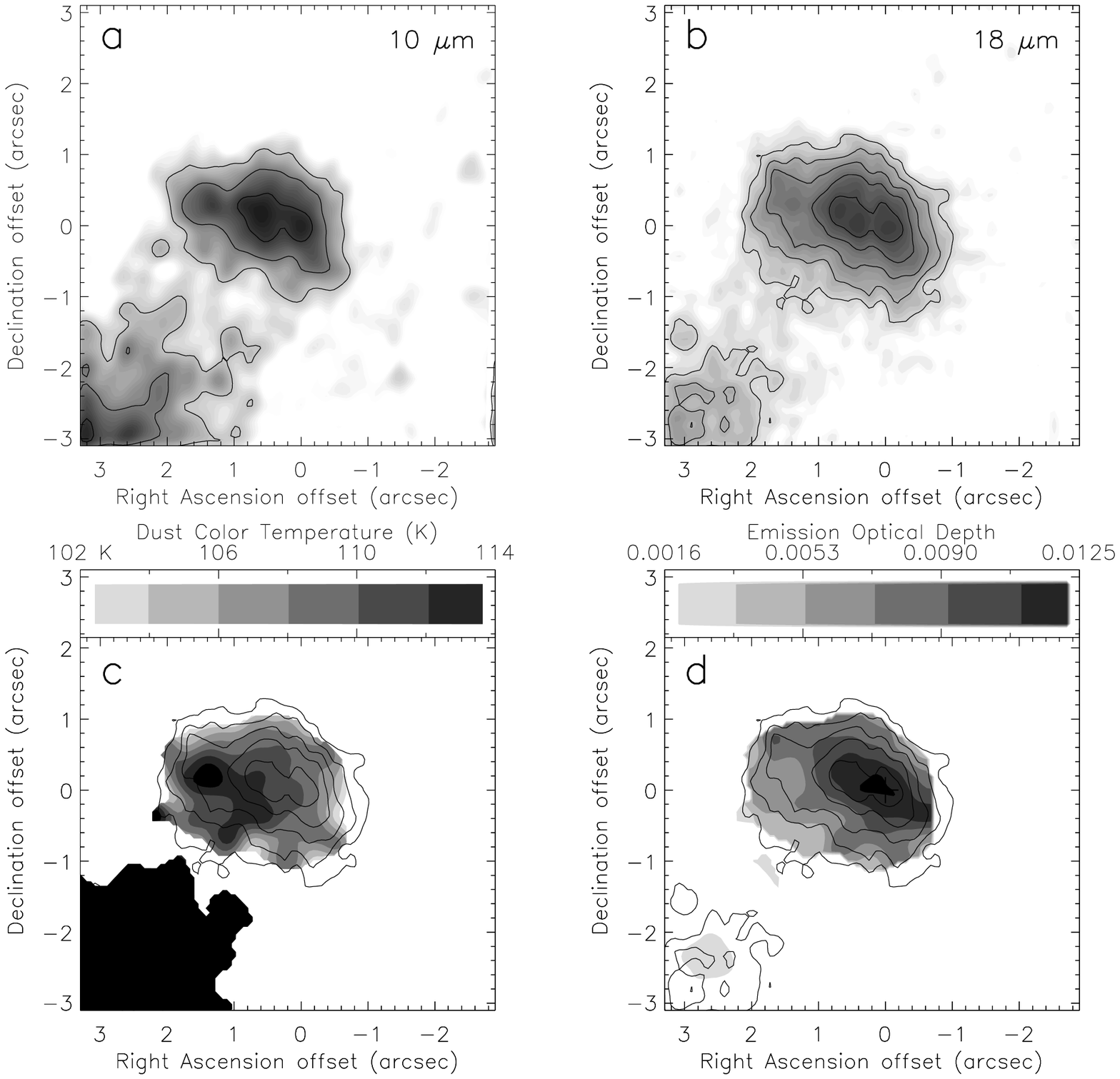}
\caption{Keck data for IRS-I 2. Panels a and b show the 
OSCIR images in grayscale at 10 and 18 {\micron}, respectively, convolved with a 
Gaussian kernel of FWHM 3 pixels. Contours are added 
for emphasis. In panel a, contour levels are 15, 40, and 60 mJy arcsec$^{-2}$. 
In panel b, contour levels are 200, 400, 600, 800, and 1000 mJy arcsec$^{-2}$.
Panel c shows the color temperature map of the region, and panel d 
shows the emission optical depth map, both with the 18 {\micron} contours 
overlaid. The origin in each panel is R.A.(J2000)=17$^h$20$^m$53\fs04 and  
Decl.(J2000)=-35$\arcdeg$46$\arcmin$58\farcs3.}
\end{figure}

The source elongation in the mid-infrared is at a position angle of $%
\thicksim $65$\arcdeg$. The methanol masers near this source \citep{E96} lie
at an angle of approximately -35$\arcdeg$, almost perpendicular to the
thermal dust elongation (Figure 5c). However, the masers are not coincident
with IRS-I 2. The measured distance from the center of the maser
distribution to the center of brightness in the 18 $\micron$ emission from
IRS-I 2 is 1.25$\arcsec$. Again, these masers lie close to the secondary
peak in the thermal ammonia emission, and may be delineating the sight of a
second hot core or embedded protostar, instead of being associated directly
with the IRS-I 2.

Looking to the color temperature maps of this source show that the hottest
part of the source is the eastern peak (Figure 7c). By overlaying the
optical depth maps with the 18 $\micron$ contours, one can see that the
optical depth distribution for the central and western sources trace the 18 $%
\micron$ contours well (Figure 7d). This implies that these two sources are
bright merely because we are seeing through more optically thin mid-infrared
emitting material. As in the case for IRS-I 1, it may be likely that the
stellar heating source does not lie at the mid-infrared peak and that the
temperature peak may be the location of the stellar source. However, there
are two reasons this argument may not be applicable in this instance. First,
unlike IRS-I 1, we cannot be sure that the temperature peak is the location
of the stellar source because there is no near-infrared component at this
location \citep{per96}. If it is hotter we would expect the energy
distribution to rise towards the near infrared, in the absence of
significant extinction. This leads to a second point, which is that the
temperature ``peak'' in this case is only 12 K hotter than the coolest parts
of the source. This is unlike IRS-I 1, where the temperature peak is
diffinitive because it is twice as hot ($\Delta $T $=60$ K) as the coolest
areas mapped. Given the fact that the source is relatively flat both in
mid-infrared flux and color temperature, IRS-I 2 may not be internally
heated. Because the temperature map shows this source to be warmest on the
side facing the ammonia peak, if there is an embedded stellar or
proto-stellar source at this location, it may be responsible for this slight
heating of IRS-I 2. This side of the molecular core would have to have less
extinction than the side facing the earth, allowing heating of IRS-I 2
without a direct view of the heating source from the earth.

There is mid-infrared emission in the southeast corner of the images of
IRS-I 2 from Keck (Figure 7a). This emission is located approximately where
we would expect the 7 mm source of \citet{ckrdh97} to be located. However,
the emission is diffuse and cut off by the edge of the array. It is most
likely just extended emission from IRS-I 1. The most recent high sensitivity
study of this source was performed by \citet{ckrmca02} and does not confirm
the detection of this source at 7 mm. They claim that the source seen in
their previous 7 mm study may have been an artifact of the data reduction
and the limited (u,v) coverage of the observations.

\subsection{IRS-I 3}

This source has low surface brightness and is hour-glass shaped in the
mid-infrared, however the northern lobe of the source is more extended in
the northern direction at 10 $\micron$ than at 18 $\micron$ (Figures 8a and
8b). \citet{KDJ99} claim that the emission drops by $\thicksim $20\% in the
area between the peaks at 20 $\micron$. Our higher resolution images show
emission drops by almost 80\% at 18 $\micron$ and 70\% at 10 $\micron$
between the peaks, so that the sources appear as separate objects in the
mid-infrared. \citet{KDJ99} speculate that this source may derive its
double-lobed shape because it is a torus or dust disk around a central star.
This seems unlikely given that the higher resolution images show elongation
in the lobes perpendicular to the plane of the speculated disk.

\begin{figure}
\plotone{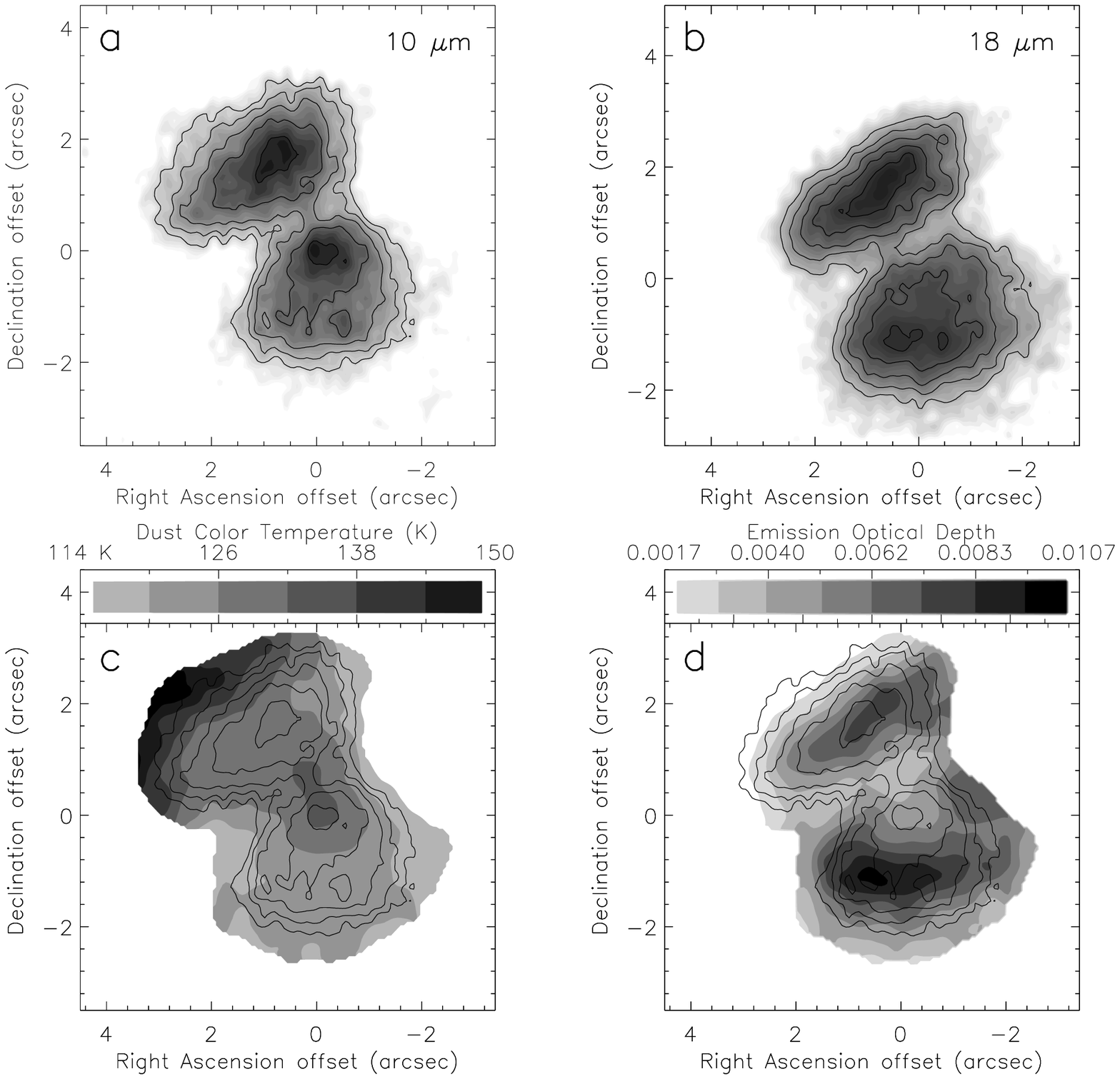}
\caption{Keck data for IRS-I 3. Panels a and b show the 
OSCIR images in grayscale at 10 and 18 {\micron}, respectively, convolved with a 
Gaussian kernel of FWHM 3 pixels. Contours are added 
for emphasis. In panel a, contour levels are 65, 85, 110, 150, and 200 mJy arcsec$^{-2}$. 
In panel b, contour levels are 600, 800, 1000, 1250, and 1600 mJy arcsec$^{-2}$.
Panel c shows the color temperature map of the region, and panel d 
shows the emission optical depth map, both with the 18 {\micron} contours 
overlaid. The origin in each panel is R.A.(J2000)=17$^h$20$^m$54\fs60 and  
Decl.(J2000)=-35$\arcdeg$47$\arcmin$02\farcs6. The 
apparent offset between panel a and b is not real, and simply a change in the 
position of the origin in the panels.}
\end{figure}

The near infrared images of \citet{per96} from this area show that the
mid-infrared emission of IRS-I 3 is coming from a cluster of ($\sim $7)
near-infrared sources with extended emission in H (Figure 9) and K. The
overall extended shape of the near-infrared emission compared to that in the
mid-infrared again strengthens our belief that the relative astrometry
between the near-infrared and mid-infrared for all these NGC 6334 I sources
is accurate. The brightest source coincident with IRS-I 3 in the H band
image of \citet{per96} appears to be located near the central temperature
peak (Figure 9b). This bright near-infrared star, which is also seen at J
and K, may be responsible for the central ionizing and heating the IRS-I 3.
Another near-infrared source is located just northeast of the northern lobe.
This is the location of the temperature maximum in the color temperature
maps, and therefore this star may be externally heating this part of the
mid-infrared source.

\begin{figure}
\plotone{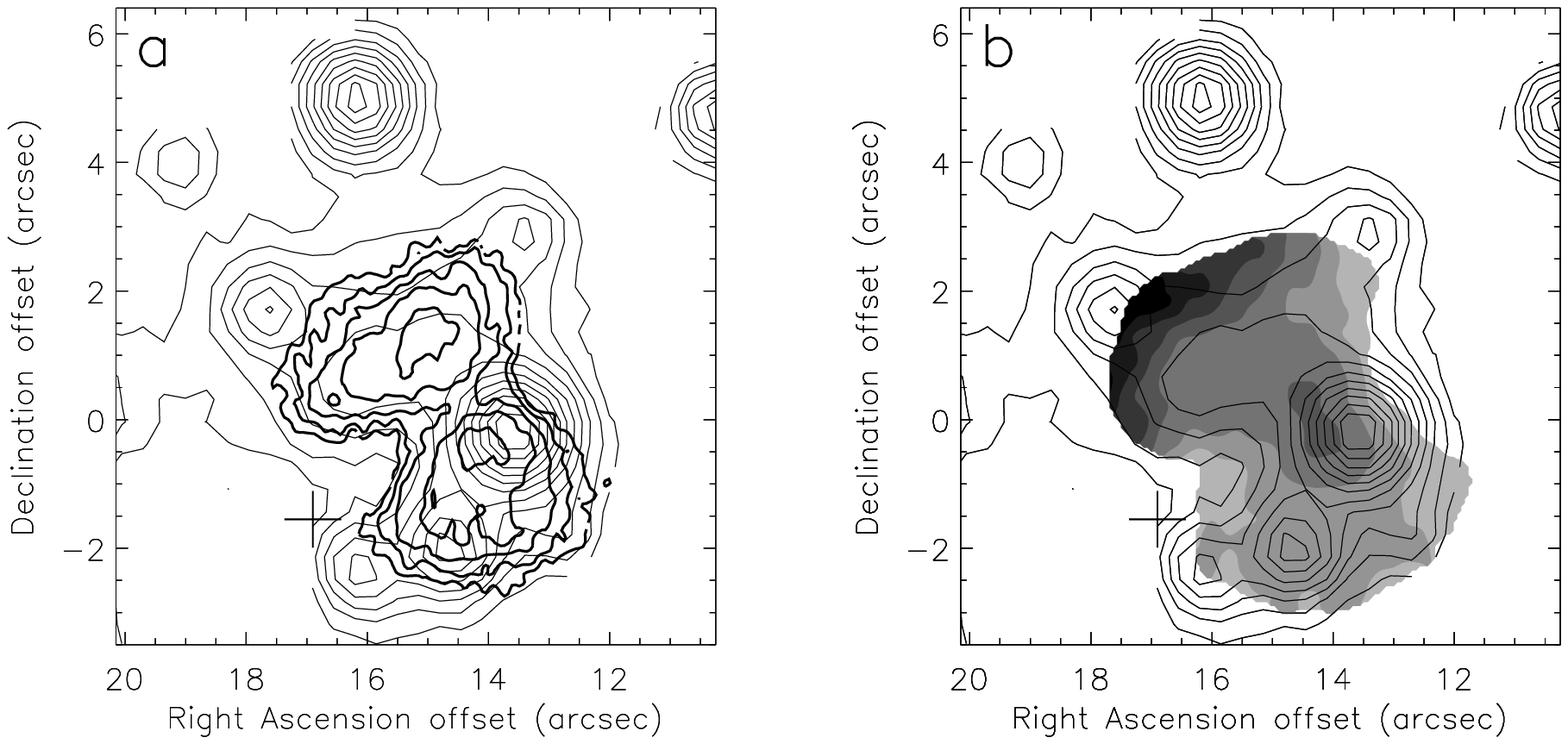}
\caption{A cluster of near infrared sources coincident with  IRS-I 3. 
Panel a shows the H band (thin contours), overlaid on the Keck 10 {\micron} extended 
emission (thick contours). Panel b shows the H band overlaid on the color temperature 
map for IRS-I 3 (filled contours). The low-spatial resolution ($\sim$8 {\arcsec}) 
2.2 {\micron } peak of Becklin and Neugebauer (1974) for IRS-I 3 is plotted here as a 
cross. This obvious discrepancy in position has lead us to a revision in the mid-infrared 
coordinates of IRS-I 3. The origin in both panels is the same as in Figure 1.}
\end{figure}

Overlaying the 20 $\micron$ image from \citet{hg83} with our 18 $\micron$
image does show good coincidence between the peaks of IRS-I 1 and IRS-I 3.
However, the coordinates quoted for this source from \citet{hg83} are
actually from the 2.2 $\micron$ observations of \citet{bn74}. This position
is offset from the position of the mid-infrared source we observe and the
near-infrared emission of \citet{per96} (Figure 9). The most likely reason
for this difference in position is because the images of \citet{bn74} were
taken with scans from a near-infrared photometer with an effective beam size
of $\sim $8$\arcsec$. The offset between the brightest near-infrared source
and the position quoted by \citet{bn74} is 3\farcs4 in right ascension and 1%
\farcs4 in declination, well within the errors they quote of 5$\arcsec$ for
right ascension and 4$\arcsec$ for declination.

Because the morphologies and peaks of IRS-I 3 are different at 10 and 18 $%
\micron$, and because of the extremely extended nature of the mid-infrared
emission, it is difficult to assign reference coordinates to the source. In
Figure 8 we arbitrarily show the 10 $\micron$ peak in the southern lobe of
IRS-I 3 as the reference position. Coordinates for this position are given
in Table 1.

\section{Conclusions}

High resolution mid-infrared observations of the central region of NGC 6334
I have revealed much about the nature and properties of the sources there.
The UCHII region NGC 6334 F is composed of two sources, IRS-I 1 and
G351.42+0.64:DPT00 2. The peak of IRS-I 1 appears to be coincident with the
peak in the radio continuum. Ammonia observations of Kraemer et al. (1999)
when registered properly with our mid-infrared data indicate that the shape
of the UCHII region is not due to a bow-shock, but instead due to
champagne-like flow from stellar source at the edge of a molecular clump.
Maser emission is concentrated at this interface between the mid-infrared
and ammonia emission, and may therefore be shock induced. There are two other strings
of masers that lie near the two peaks in the ammonia emission and may be
delineating the sites of hot molecular cores that are too young and/or
embedded to be seen yet in the mid-infrared.

The mid-infrared emission from IRS-I 1 seems to be coming from an arc of
dust at the interface between the molecular ammonia clump and the UCHII
region, and may be material swept up by the expanding shock front of the
UCHII\ region. The color temperature peaks at a location interior to this
mid-infrared arc, coincident with a near infrared source IRS1E. This source
may be the stellar source responsible for the ionization and heating of the
NGC\ 6334 F region. Two other near infrared sources, IRS1W and IRS1SW, lie
in the northern and southern parts of the mid-infrared arc and are
associated with the majority of the masers in the region. There is no
temperature peak at these locations, so the near infrared emission may just
be reflected or shock excited emission. A fourth near infrared source
(IRS1SE) seems to simply be reflected emission off the UCHII\ region tail.

G351.42+0.64:DPT00 2 appears to be a clump of dust, perhaps swept up by the
shock front of UCHII region. It displays a steep temperature gradient
towards the color temperature peak. It also shows some signs of ionized
emission in the higher resolution radio continuum images, but only on the
hotter, southern side. For these reasons it may be that DPT00 2 has no
central heating source but is simply heated and ionized by the same source
heating and ionizing IRS-I 1 (IRS1E) .

IRS-I 2 was believed to be associated with the a linear structure of
methanol masers and perhaps delineating a circumstellar disk. However, the
thermal dust emission is elongated at a different position angle to the
position angle of the maser distribution. The low-surface brightness, smooth
color temperature distribution, and lack of a near-infrared component may
indicate that there is no internal stellar source here at all. Furthermore,
the masers are offset from the mid-infrared peak and could be associated
with the secondary peak in the ammonia distribution.

Lower resolution mid-infrared images of IRS-I 3 showed it to be a double
peaked source. However, the high resolution images presented here show that
it has a complex and peculiar morphology. We find, using the near infrared
data of Persi et al. (1996), that the large and extended mid-infrared
sources are extended dust emission from a cluster of stellar sources seen in
the near infrared. These stellar sources have directly influenced the
morphology in the mid-infrared, and the structure of the source as seen in
the color temperature map.

The reality of the 7 mm source in this region has been seriously called into
question by the non-detection in the mid-infrared, and has recently been
discovered to be an artifact of data reduction by follow-up observations of
the original authors.

\acknowledgments The authors would like to thank NASA's Florida Space Grant
Consortium for their financial support of the first author at the time of
observations. Data presented herein were obtained at the W.M. Keck
Observatory, which is operated as a scientific partnership among the
California Institute of Technology, the University of California and the
National Aeronautics and Space Administration. The Observatory was made
possible by the generous financial support of the W.M. Keck Foundation. The
authors wish to recognize and acknowledge the very significant cultural role
and reverence that the summit of Mauna Kea has always had within the
indigenous Hawaiian community. We are most fortunate to have the opportunity
to conduct observations from this mountain.

\clearpage

\begin{deluxetable}{lccccccc}
\rotate
\scriptsize
\tablewidth{0pt}
\tablecaption{Properties of Sources in NGC 6444 I \label{tbl-1}}
\tablehead{
\colhead{Source} & \colhead{10 {\micron} Flux}   & \colhead{18 {\micron} Flux}  & \colhead{Offset \tablenotemark{\natural} ($\Delta\alpha$, $\Delta\delta$)} & \colhead{Right Ascension \tablenotemark{\natural} } & \colhead{Declination \tablenotemark{\natural} }&\colhead{Luminosity\tablenotemark{\star}} &\colhead{ZAMS}\\
\colhead{} & \colhead{ Density\tablenotemark{\dagger} (Jy)}   & \colhead{Density\tablenotemark{\dagger} (Jy)} & \colhead{(arcsec)} &\colhead{(J2000)} &\colhead{(J2000)} &\colhead{(L$_{\sun}$)} &\colhead{Spectral Type\tablenotemark{\star}}
} 

\startdata
IRS-I 1 &78.63$\pm$5.42 &222.70$\pm$21.60 &(0,0) &17 20 53.44 &-35 47 02.2 &3285 &B2\nl
IRS-I 2\tablenotemark{\ddagger} &0.13$\pm$0.01 &3.41$\pm$0.33 &(-4.8,+3.9) &17 20 53.04 &-35 46 58.3 &67 &B9\nl
IRS-I 3 (all) &2.22$\pm$0.15 &22.96$\pm$2.23 &(+14.1,-0.4) &17 20 54.60 &-35 47 02.6 &326 &B7\nl
...north lobe &1.09$\pm$0.07 &8.57$\pm$0.83 &\nodata &\nodata &\nodata &117 &B9\nl
...south lobe &1.24$\pm$0.08 &12.87$\pm$1.25 &\nodata &\nodata &\nodata &183 &B8\nl
DPT00 2 &11.37$\pm$0.77 &68.65$\pm$6.66 &(+4.0,+1.6) &17 20 53.77 &-35 47 00.6 &919 &B4\nl
\enddata

\tablenotetext{\dagger}  {Color corrected flux densities. The quoted errors in the measurements are the absolute photometric accuracy for the night (the dominant source of error), which was calculated to be 6.8\% at 10 {\micron} and 9.7\% at 18 {\micron}. All extended emission in the frames is included for all sources, and may lead to the small differences between the values quoted here and that of De Buizer, Pi\~{n}a, \& Telesco (2000) .}
\tablenotetext{\natural}  {Refer to the text and Figures 7 and 8 for locations of these coordinates for the extended sources IRS-I 2 and IRS-I 3.}
\tablenotetext{\star}  {Luminosities and spectral types are given for all objects for the sake of completeness. Some sources may not actually have stellar cores, or may be heated by more than one star (see text).}
\tablenotetext{\ddagger}  {Flux densities for IRS-I 2 are much different than those quoted in De Buizer, Pi\~{n}a, \& Telesco (2000). This is most likely due to the fact that the CTIO 4-m images yielded very low signal-to-noise for IRS-I 2.}
\end{deluxetable}

\end{document}